\documentclass[12pt]{article}
\usepackage{subfloat}
\pdfoutput = 1
\usepackage{graphics}
\usepackage{graphicx} %Include figure filesusepackage{graphicx} %Include figure files
\usepackage{color}
\usepackage{multirow}
\usepackage{hhline}
\begin{document}
\date{Today}
%%%%%%%%%%%%%%%%%%%%
\title{{\bf{\Large Conductivity of holographic superconductors in Born-Infeld electrodynamics }}}
%%%%%%%%%%%%%%%%%%%%

\author{
{\bf {\normalsize Debabrata Ghorai}$^{a}$
\thanks{debanuphy123@gmail.com, debabrataghorai@bose.res.in}},\,
{\bf {\normalsize Sunandan Gangopadhyay}$^{a}
$\thanks{sunandan.gangopadhyay@gmail.com, sunandan.gangopadhyay@bose.res.in}}\\
$^{a}$ {\normalsize Department of Theoretical Sciences,}\\{\normalsize S.N. Bose National Centre for Basic Sciences,}\\{\normalsize JD Block, 
Sector III, Salt Lake, Kolkata 700098, India}\\[0.6 cm]
}
\date{}

\maketitle

\begin{abstract}
{\noindent In this paper, we have analytically computed the conductivity of holographic superconductors in the framework of Born-Infeld electrodynamics taking into account the backreaction of the matter fields on the bulk spacetime metric. The effect of the Born-Infeld electrodynamics is incorporated in the metric. The band gap energy is found to be corrected by the backreaction and Born-Infeld parameters. The conductivity expression is then compared with that obtained from a self consistent approach.  }
\end{abstract}
\vskip 1cm

%%%%%%%%%%%%%%%%%%%%%%%%%%%%%%%%% Introduction %%%%%%%%%%%%%%%%%%%%%%%%%%%%%%%%%%%%%%%%%%%%%%%%%%%%
\section{Introduction}
The AdS/CFT duality has been an important theoretical input to study the physics of strongly coupled system \cite{adscft1}-\cite{nw2}. The main focus in this area has been to construct gravitational duals of physical phenomena exhibiting strong coupling.

 Holographic superconductors have been an important class of gravitation duals which have been studied extensively in the recent past \cite{hs6}-\cite{caih}. These models have been found to reproduce some of the properties of strongly coupled superconductors. The theoretical models consisting of a AdS black hole in the bulk with a charged scalar field coupled to the Maxwell field has been found to admit the formation of a scalar hair below a certain critical temperature. The mechanism involved in the formation of this hair involves the spontaneous breakdown of a local $U(1)$ symmetry near the black hole horizon \cite{hs1},\cite{hs2}.
 
There has been a lot of work investigating the effects of Born-Infeld (BI) electrodynamics on holographic superconductors \cite{hs19}-\cite{dg4}. The importance of such work have been to study the effect of non-linear electrodynamics on holographic superconductors. Further, the choice of BI electrodynamics have been made since this is the only non-linear theory of electrodynamics which enjoys the duality symmetry. However, an analytic computation of conductivity of holographic superconductors in the framework of BI electrodynamics has so far been missing in the literature. In this paper, we proceed to investigate the effects of BI electrodynamics on the conductivity of these systems analytically. The computation of the conductivity analytically incorporating the non-linear effects of BI electrodynamics is important in its own right as it would give information about the dependence of the band gap energy on the non-linear effects coming from BI electrodynamics. We incorporate the effects of the BI parameter in the spacetime metric and also take backreaction into account. We have then calculated the band gap energy from the conductivity expression. We have shown that the band gap energy increases with increase in parameter $b$.

This paper is organized as follows. In section 2, the basic formalism for the holographic superconductors coupled to BI electrodynamics is presented. In section 3, we calculate the conductivity upto first order in $b$. Section 4 contains the concluding remarks.

%%%%%%%%%%%%%%%%%%%%%%%%%%%%%%%%%%%%%%%%%%%%%%%%%%%%%%%%%%%%%%%%%%%%%%%%%%%%%%%%%%%%%%%%%%%

%%%%%%%%%%%%%%%%%%%%%%%%%%%%%  Section 2     %%%%%%%%%%%%%%%%%%%%%%%%%%%%%%%%%%%%%%%%%%%%%%

\section{Basic formalism }
In this section, we set up the basic formalism and notations which shall be required for subsequent discussion. 
In $3+1$-dimensions, the action for the model of a holographic superconductor in the framework of Born-Infeld electrodynamics consists a complex scalar field coupled to a $U(1)$ gauge field in anti-de Sitter black hole spacetime 
\begin{eqnarray}
S=\int d^{4}x \sqrt{-g} \left[ \frac{1}{2 \kappa^2} \left( R -2\Lambda \right) +\frac{1}{b}\left(1-\sqrt{1+\frac{b}{2} F^{\mu \nu} F_{\mu \nu}}\right) -(D_{\mu}\psi)^{*} D^{\mu}\psi-m^2 \psi^{*}\psi \right]
\label{tg1}
\end{eqnarray}
where $F_{\mu \nu}=\partial_{\mu}A_{\nu}-\partial_{\nu}A_{\mu}$; ($\mu,\nu=t,r,x,y$), $D_{\mu}\psi=\partial_{\mu}\psi-iqA_{\mu}\psi$, $\Lambda=-\frac{3}{L^2}$ is the cosmological constant, $\kappa^2 = 8\pi G $, $G$ being the Newton's universal gravitational constant,  $b$ is Born-Infeld parameter, $A_{\mu}$ and $ \psi $ represent the gauge and scalar fields. 

In the presence of backreaction, the plane-symmetric black hole metric takes the form
\begin{eqnarray}
ds^2=-f(r)e^{-\chi (r)} dt^2+\frac{1}{f(r)}dr^2+ r^2 (dx^2 + dy^2)~.
\label{tg2}
\end{eqnarray}
\noindent Making the ansatz for the gauge field and the scalar field as \cite{hs3}
\begin{eqnarray}
A_{\mu} = (\phi(r),0,0,0)~,~\psi=\psi(r)
\label{tg3}
\end{eqnarray}
leads to the following equations of motion for the metric, the gauge and matter fields 
\begin{eqnarray}
&& f^{\prime}(r) + \frac{f(r)}{r} - \frac{3r}{L^2} +  \kappa^2 r \nonumber\\
&\times & \left[f(r)\psi^{\prime}(r)^2 + \frac{q^2 \phi^2(r) \psi^2(r) e^{\chi(r)}}{f(r)} + m^2 \psi^2(r)  + \frac{1}{b}\left((1- b e^{\chi(r)}\phi^{\prime}(r)^{2})^{-\frac{1}{2}} -1\right)\right]=0,\nonumber \\
\label{e01}
\end{eqnarray}
\begin{eqnarray}
\chi^{\prime}(r) + 2 \kappa^2 r \left(\psi^{\prime}(r)^2 + \frac{q^2 \phi^2(r) \psi^2(r) e^{\chi(r)}}{f(r)^2}\right) = 0,
\label{e02}
\end{eqnarray}
\begin{eqnarray}
\phi^{\prime \prime}(r) + \left(\frac{2}{r}+ \frac{\chi^{\prime} (r)}{2}\right) \phi^{\prime}(r) - \frac{2}{r} b e^{\chi(r)}\phi^{\prime}(r)^{3} - \frac{2 q^2 \phi(r) \psi^{2}(r)}{f(r)}(1 - b e^{\chi(r)} \phi^{\prime}(r)^{2})^\frac{3}{2} = 0,
\label{e1}
\end{eqnarray}
\begin{eqnarray}
\psi^{\prime \prime}(r) + \left(\frac{2}{r}- \frac{\chi^{\prime} (r)}{2} + \frac{f^{\prime}(r)}{f(r)}\right)\psi^{\prime}(r) + \left(\frac{q^2 \phi^{2}(r) e^{\chi(r)}}{f(r)^2}- \frac{m^{2}}{f(r)}\right)\psi(r) = 0
\label{e03}
\end{eqnarray}
where prime denotes derivative with respect to $r$. We can set $q=1$ and $L=1$ without any loss of generality \cite{yao}, \cite{chen}. The conditions $\phi(r_+)=0$ and $\psi(r_{+})$ to be finite imposes the regularity of the fields at the horizon. 

The fields near the boundary of the bulk obey \cite{hs8}
\begin{eqnarray}
\label{bound1}
\phi(r)&=&\mu-\frac{\rho}{r}~,\\
\psi(r)&=&\frac{\psi_{-}}{r^{\Delta_{-}}}
+\frac{\psi_{+}}{r^{\Delta_{+}}}
\label{bound2}
\end{eqnarray}
where 
\begin{eqnarray}
 \Delta_{\pm} = \frac{3\pm\sqrt{9+4m^2 L^2}}{2}~
\label{del}
\end{eqnarray}
are the conformal weights of the conformal field theory living on the boundary. The interpretation of the parameters $\mu$ and $\rho$ is given by the gauge/gravity dictionary. They are interpreted as the chemical potential and charge density of the conformal field theory on the boundary. For the choice $\psi_{+}=0$, $\psi_{-}$ is interpreted as the dual of the expectation value of the condensation operator $\mathcal{O}_{\Delta}$ in the boundary. 

Under changing the coordinate from $r$ to $z=\frac{r_+}{r}$, the field eq.(s) (\ref{e01})-(\ref{e03}) look like
\begin{eqnarray}
&& f^{\prime}(z) - \frac{f(z)}{z} + \frac{3r^2_{+}}{z^3} - \frac{ \kappa^2 r^2_{+}}{z^3} \nonumber\\ &\times& \left[\frac{z^4}{r^2_{+}} f(z)\psi^{\prime}(z)^2 + \frac{\phi^2(z) \psi^2(z) e^{\chi(z)}}{f(z)} + m^2 \psi^2(z)  + \frac{1}{b}\left((1- \frac{b z^4}{r^2_{+}}e^{\chi(z)} \phi^{\prime}(z)^{2})^{-\frac{1}{2}} -1\right)\right]=0,\nonumber\\
\label{e1a}
\end{eqnarray}
\begin{eqnarray}
\chi^{\prime}(z) - \frac{2 \kappa^2 r^2_{+}}{z^3}\left(\frac{z^4}{r^2_{+}}\psi^{\prime}(z)^2 + \frac{\phi^2(z) \psi^2(z) e^{\chi(z)}}{f(z)^2}\right) = 0,
\label{e1ab}
\end{eqnarray}
\begin{eqnarray}
\phi^{\prime \prime}(z) + \frac{\chi^{\prime}(z)}{2} \phi^{\prime}(z) + \frac{2}{r^2_{+}} b e^{\chi(z)}\phi^{\prime}(z)^{3} z^3 - \frac{2r^2_{+} \phi(z) \psi^{2}(z)}{f(z) z^4}\left(1 -\frac{b z^4 e^{\chi(z)}}{r^2_{+}} \phi^{\prime}(z)^{2}\right)^\frac{3}{2} = 0, \nonumber\\
\label{e1aa}
\end{eqnarray}
\begin{eqnarray}
\psi^{\prime \prime}(z) + \left(\frac{f^{\prime}(z)}{f(z)} - \frac{\chi^{\prime}(z)}{2}\right)\psi^{\prime}(z) + \frac{r^2_{+}}{z^4} \left(\frac{\phi^{2}(z) e^{\chi(z)}}{f(z)^2}- \frac{m^{2}}{f(z)}\right)\psi(z) =0 
\label{psiz}
\end{eqnarray}
where prime denotes derivative with respect to $z$. 
%In general the matter field takes the form as
%\begin{eqnarray}
%\psi(r) = \frac{\langle \mathcal{O}_{\Delta}\rangle}{\sqrt{2} r^{\Delta}} F(r)
%\end{eqnarray}
%where $F(\infty) = 1$ and $ \mathcal{O}_{\Delta}$ is the condensation operator.
The regularity condition $\phi(r_+)$ becomes $\phi(z=1)= 0$. In the rest of our work, we set $m^2 = -2$. This leads to two possible values of $\Delta$ from eq.(\ref{del}), namely, $\Delta_{+}=2$ and $\Delta_{-}=1$.

%\textbf{In general one can investigate both cases but we are mainly focusing the $\Delta_{-}=1$ case.} \\
We now proceed to solve the equation for the metric (\ref{e1a}) taking into account the effect of the backreaction and the BI parameter $b$.
At $T=T_c$, the matter field vanishes, that is $\psi(z) =0$.
Hence eq.(\ref{e1ab}) reduces to 
\begin{eqnarray}
\label{bk1}
\chi^{\prime}(z)= 0 ~~~\Rightarrow \chi (z) =constant~.
\label{bk2}
\end{eqnarray}
Now near the boundary of the bulk, we can set $e^{-\chi(r\rightarrow\infty)} \rightarrow 1 $, i.e. $\chi(r\rightarrow\infty)=0$
which in turn implies $\chi(z=0)=0$. This yields $\chi(z)=0$ from eq.(\ref{bk2}). Therefore, the gauge field equation (\ref{e1aa}) reduces to 
\begin{eqnarray}
\label{metric1}
\phi^{\prime \prime}(z)  + \frac{2bz^3}{r^2_{+(c)}} \phi^{\prime}(z)^3=0
\end{eqnarray}
where $r_{+(c)}$ is the horizon radius at $T=T_c$.  The solution of this equation for $\phi(z)$  upto $\mathcal{O}(b)$ subject to the boundary condition (\ref{bound1}) reads \cite{sgdc1},\cite{dg1}
\begin{eqnarray}
\phi(z)= \lambda r_{+(c)}\left\{(1-z)-\frac{b\lambda^2}{10}(1-z^5)\right\}
\label{chs20}
\end{eqnarray}
where 
\begin{eqnarray}
\lambda = \frac{\rho}{r^{2}_{+(c)}}.
\label{lamr}
\end{eqnarray} 
With these solutions in hand, we now proceed to solve the equation for the metric.
The metric equation keeping terms upto first order in the Born-Infeld parameter now reads
\begin{eqnarray}
f^{\prime}(z) - \frac{f(z)}{z} +\frac{3r^{2}_{+(c)}}{z^3} -\kappa^2  \left(\frac{1}{2}\phi^{\prime 2}(z) z +\frac{3b z^5}{8r^{2}_{+(c)}} \phi^{\prime 4}(z) \right) = 0 ~.
\label{chs19}
\end{eqnarray}
Substituting the solution of $\phi(z)$ in the above equation, we obtain the metric equation upto $\mathcal{O}(b)$
\begin{eqnarray}
f^{\prime} - \frac{f(z)}{z} + \frac{3r^{2}_{+(c)}}{z^3} - \frac{r^2_{+(c)}\kappa^2 \lambda^2}{2}\left(z - \frac{b}{4}\lambda^2 z^5 \right)=0~.
\end{eqnarray}
Solving this equation and imposing the condition $f(z=1)=0$ to determine the integration constant yields
\begin{eqnarray}
f(z)= \frac{r_{+(c)}}{z^2} g(z) \equiv \frac{r_{+(c)}}{z^2} \left[ g_{0}(z) + g_{1}(z)\right]
\label{chs21}
\end{eqnarray}
where 
\begin{eqnarray}
g_{0}(z)=1-z^3 ~~~;~~~ g_{1}(z)=\frac{\kappa^2\lambda^2}{2}\left\{z^4 -z^3 -\frac{b\lambda^2}{20}(z^8-z^3)\right\}~.
\label{chs22}
\end{eqnarray}
The above form of the metric includes the effects of backreaction as well as the BI electrodynamics upto first order in the BI parameter $b$. The Hawking temperature of this black hole spacetime reads
\begin{eqnarray}
T = \frac{f^{\prime}(r_{+})}{4\pi} = -\frac{f^{\prime}(z=1)}{4\pi r_{+}} = \frac{3r_{+}}{4\pi} \left[ 1- \frac{\kappa^2 \lambda^2}{6}\left(1-\frac{b\lambda^2}{4}\right)\right]
\label{tcm}
\end{eqnarray} 
which is interpreted as the temperature of the dual field theory at the boundary. Substituting eq.(\ref{lamr}) in the above equation, we obtain the relation between the critical temperature and the charge density to be
\begin{eqnarray}
T_c = \frac{3}{4\pi} \left[ 1- \frac{\kappa^2_{i} \lambda^2_{i-1}}{6}\left(1-\frac{b(\lambda^2 |_{b=0})}{4}\right)\right]\sqrt{\frac{\rho}{\lambda}} \equiv \xi\sqrt{\rho}~.
\label{tcf}
\end{eqnarray}  
Note that $\kappa^2_{i} \lambda^2 =\kappa^2_{i}(\lambda^2|_{i-1}) +\mathcal{O}(\kappa^4)$ and $b\lambda^2 = b(\lambda^2 |_{b=0}) +\mathcal{O}(b^2) $ in the above equation.
The procedure to estimate the values of $\xi$ is as follows \cite{dg1}. The matter field equation (\ref{psiz}) is first solved near $T_c$. At $T\rightarrow T_c$ (but $T\neq T_c$), the equation for the matter field (\ref{psiz}) reduces to \begin{eqnarray}
\psi''(z)+ \left(\frac{g'(z)}{g(z)}-\frac{2}{z}\right)\psi'(z)+ \left( \frac{\phi^{2} (z)}{g^{2} (z) r^{2}_{+(c)}} + \frac{2}{g(z) z^2}\right)\psi(z)=0
\label{e001}
\end{eqnarray}
where $\phi(z)$ now corresponds to the solution (\ref{chs20}).  
Near the boundary, we define for $\Delta=1$ \cite{siop}
\begin{eqnarray}
\psi(z)=\frac{\langle\mathcal{O}_{1}\rangle}{\sqrt{2} r_{+(c)}} z F(z)
\label{sol1}
\end{eqnarray}
where $F(z)$ is a trial function with $F(0)=1, F'(0)=0 $ and $\langle\mathcal{O}_{1}\rangle $ is the condensation operator.
Substituting this form of $\psi(z)$ in eq.(\ref{e001}), we obtain
\begin{eqnarray}
F''(z) &+& \left\{\frac{g'(z)}{g(z)} \right\}F'(z) +\left\{  \left(\frac{g'(z)}{g(z)}-\frac{2}{z}\right)\frac{1}{z}+\frac{2}{g(z) z^2} \right\}F(z) \nonumber \\
&+& \frac{\lambda^2}{g^{2}(z)}\left\{ (1-z)^2 -\frac{b(\lambda^2|_{b=0})}{5}(1-z)(1-z^5)\right\}F(z)=0~.
\label{eq5b}
\end{eqnarray}
Recasting the above equation in the Sturm-Liouville form gives
\begin{eqnarray}
\frac{d}{dz}\left\{p(z)F'(z)\right\}+q(z)F(z)+\lambda^2 r(z)F(z)=0
\label{sturm}
\end{eqnarray}
with 
\begin{eqnarray}
p(z)&=& g(z),\nonumber\\
q(z)&=& g(z)\left\{ \left(\frac{g'(z)}{g(z)}-\frac{2}{z} \right)\frac{1}{z}+\frac{2}{g(z) z^2} \right\}, \nonumber\\
r(z)&=&\frac{1}{g(z)} \left\{ (1-z)^2 -\frac{b(\lambda^2|_{b=0})}{5}(1-z)(1-z^{5})\right\}. 
\label{i1}
\end{eqnarray}
With these identifications, one can write down an equation for the eigenvalue $\lambda^2$ which minimizes the expression 
\begin{eqnarray}
\lambda^2 &=& \frac{\int_0^1 dz\ \{p(z)[F'(z)]^2 - q(z)[F(z)]^2 \} }
{\int_0^1 dz \ r(z)[F(z)]^2}~.
\label{eq5abc}
\end{eqnarray}
We may now consider the following trial function for the estimation of $\lambda^{2}$
\begin{eqnarray}
F= F_{\alpha} (z) \equiv 1 - \alpha z^2 ~.
\label{eq50}
\end{eqnarray}
This function satisfies the conditions $F(0)=1$ and $F'(0)=0$. 
Substituting eq.(s)(\ref{i1}) and (\ref{eq50}) in eq.(\ref{eq5abc}) and setting the backreaction parameter $\kappa=0$ and the Born-Infeld parameter $b=0$ yields
\cite{siop}
\begin{eqnarray}
\lambda_{\alpha}^2 = \frac{2(3-3\alpha+5{\alpha}^2 )}{(2\sqrt{3}\pi -6\ln3)+4(\sqrt{3}\pi+3\ln3-9)\alpha + (12\ln3-13)
{\alpha}^2} ~.
\label{est2}
\end{eqnarray}
The minimum value of $\lambda_{\alpha}^2 =1.2683$ and occurs at $\alpha \approx 0.2389$. This in turn gives the value of $\xi$. Eq.(\ref{tcf}) now gives the critical temperature  
\begin{eqnarray}
T_c =\frac{3}{4\pi\sqrt{\lambda|_{\tilde\alpha=0.2389}}}\sqrt{\rho} \approx 0.2250\sqrt{\rho} 
\label{eqTc}
\end{eqnarray}
which is in very good agreement with the numerical result
$T_c = 0.226\sqrt{\rho}$ \cite{hs6}.

To include the effects of the Born-Infeld and back reaction parameters $b$
and $\kappa$, we proceed as follows. We set different values of $b$ and $\kappa$ and rerun the above analysis to get the value of $\lambda^2$. This in turn gives the relation between the critical temperature and the charge density for different values of $\kappa$ and $b$.
%For $b=0.0$ and $\kappa=0.1$
%\begin{eqnarray}
%\lambda_{\alpha}^2 = \frac{0.498943 - 0.498098 \alpha + 0.831974 \alpha^2}{ 0.357672 - 4.27217 \alpha + 0.0387435 \alpha^2}.
%\label{est2}
%\end{eqnarray}
%The minimum value of $\lambda_{\alpha}^2 = 1.2661$ and occurs at $\alpha \approx 0.238201$. This in turn gives the value of $\xi$. Eq.(\ref{tcf}) now gives the critical temperature  
%\begin{eqnarray}
%T_c =\frac{3}{4\pi\sqrt{\lambda|_{\tilde\alpha=0.23820}}}\left[1-\frac{(0.1)^2 (1.26832)}{6} \right] \sqrt{\rho} \approx 0.2246\sqrt{\rho} 
%\label{eqTc}
%\end{eqnarray} 

Setting $b=0.1$ and $\kappa=0.1$ in eq.(s)(\ref{i1}) and (\ref{chs22}) and using them in eq.(\ref{eq5abc}) along with eq.(\ref{eq50}), we obtain the value of $\lambda^2$ in terms of $\alpha$ to be 
\begin{eqnarray}
\lambda_{\alpha}^2 = \frac{0.498954 - 0.498131 \alpha + 0.832002 \alpha^2}{ 0.344004 - 0.0825083 \alpha + 0.0141889 \alpha^2}.
\label{est2}
\end{eqnarray}
The minimum value of $\lambda_{\alpha}^2 = 1.31478$ and occurs at $\alpha \approx 0.23954$. This in turn gives the value of $\xi$. Eq.(\ref{tcf}) now gives the critical temperature  
\begin{eqnarray}
T_c \approx 0.2225\sqrt{\rho}~.
\label{eqTc}
\end{eqnarray} 
In Table \ref{t1}, we present the values of $\lambda^2$ for different values of $\kappa$ and $b$. These results shall be used in the next section to calculate the band gap energy for different values $\kappa$ and $b$.
%\frac{3}{4\pi\sqrt{\lambda|_{\tilde\alpha=0.23954}}}\left[1-\frac{(0.1)^2 (1.26832)}{6}\left(1- \frac{(0.1)1.26607}{4} \right) \right] \sqrt{\rho}

\begin{table}[h!]
\caption{The values of $\lambda^2 $ for different values of $\kappa$ and $b$.}
\centering
\begin{tabular}{|c| c| c| c| }
\hline
$\lambda^2 $ & $\kappa =0.1$ & $\kappa =0.2$ & $\kappa = 0.3$  \\
\hline
$b=0.0$ ~& ~1.2661~ & ~1.2593~ & ~1.2481~   \\
\hline
$b=0.1$ & 1.3148 & 1.3076 & 1.2956  \\
\hline
$b=0.2$ & 1.3674 & 1.3597 & 1.3471  \\
\hline
$b=0.3$ & 1.4244 & 1.4162 & 1.4028 \\
\hline
\end{tabular}
\label{t1}
\end{table}

\section{Computation of conductivity} 
In this section, we proceed to study the conductivity as a function of frequency, that is optical conductivity. For simplicity, we look at the conductivity along the $x$-direction. By the gauge/gravity duality, the fluctuations in the Maxwell field in the bulk gives rise to the conductivity.

Making the ansatz $A_{\mu} = (0,0,\varphi(r,t), 0)$ with $\varphi(r,t) = A(r)e^{-i\omega t}$ and neglecting terms of $\mathcal{O}(b^2)$ and $ \mathcal{O}(\omega^2 b)$ leads to the following equation of motion for $A(r)$
\begin{eqnarray}
A^{\prime\prime}(r) +\frac{f^{\prime}(r)A^{\prime}(r)}{f(r)}\left\{  1+ \frac{b}{r^2}f(r) A^{\prime 2}(r) e^{-2i\omega t}\right\} -\frac{b e^{-2i\omega t}}{2r^2}A^{\prime 3}(r)\left(f^{\prime}(r) - \frac{2f(r)}{r}\right) \nonumber \\
+\left[\frac{\omega^2}{f^{2}(r)} -\frac{2\psi^{2}(r)}{f(r)}\left(1+ \frac{3b}{2r^2}f(r)A^{\prime 2}(r) e^{-2i\omega t} \right)\right]A(r) = 0~.
\label{chs3}
\end{eqnarray}
This equation is very difficult to solve analytically. However, in principle we can employ a perturbative approach to tackle this equation.

To make progress, we start by neglecting the non-linear terms in the above equation \footnote{This is done to carry out the analysis analytically}. This reads 
\begin{eqnarray}
A^{\prime\prime}(r) +\frac{f^{\prime}(r)A^{\prime}(r)}{f(r)} +
\left[\frac{\omega^2}{f^{2}(r)} -\frac{2\psi^{2}(r)}{f(r)}\right]A(r) = 0 ~.
\label{chs23}
\end{eqnarray}
Note that the effect of the BI parameter is contained in the metric. The perturbative technique involves solving this equation and then replacing this solution in the $\mathcal{O}(b)$ terms in eq.(\ref{chs3}) and solving the equation once again.

We now move to tortoise coordinate which is defined by
\begin{eqnarray}
r_{*} &=& \int \frac{dr}{f(r)} =-\frac{1}{r_{+}} \int \frac{dz}{g_{0}(z)+g_{1}(z)} \approx -\frac{1}{r_{+}}\left\{\int \frac{dz}{g_{0}(z)} -\int \frac{g_{1}(z)}{g_{0}^{2}(z)}dz\right\} \nonumber \\
&=& ln(1-z)^{\frac{1}{3r_+}\left\{1+\frac{\kappa^2\lambda^2}{6}\left(1-\frac{1}{4}b\lambda^2\right)\right\}} + ln(1+z+z^2)^{-\frac{1}{6r_+}\left\{1+\frac{\kappa^2\lambda^2}{6}\left(1+\frac{13}{20}b\lambda^2\right)\right\}} \nonumber \\
&+&\frac{\kappa^2 \lambda^4 b}{120r_+}(1-z^3)-\frac{\kappa^2\lambda^2 (z+\frac{b\lambda^2}{20})}{6r_{+}(1+z+z^2)} -\frac{1}{\sqrt{3}r_+}\left\{1-\frac{\kappa^2\lambda^2}{2}+\frac{\kappa^2\lambda^4 b}{120} \right\} tan^{-1}\frac{\sqrt{3}z}{2+z}~. \nonumber \\
\label{chs24}
\end{eqnarray}
The integration constant has been calculated from the condition $r_{*}(z)=0$ at $z=0$.
The wave equation (\ref{chs23}) in tortoise coordinate reads
\begin{eqnarray}
\frac{d^{2}A}{dr_{*}^{2}} + \left[\omega^2 -V \right]A = 0
\label{chs25}
\end{eqnarray}
where
\begin{eqnarray}
V= 2 \psi^2 f ~.
\label{chs26}
\end{eqnarray}
We now employ a trick to solve this equation. We first solve this equation for $V=0$ which implies that we solve only the $\omega$-dependent part of the equation. The solution reads
\begin{eqnarray}
A \sim e^{-i\omega r_{*}} \sim (1-z)^{-\frac{i\omega}{3r_+}\left\{1+\frac{\kappa^2\lambda^2}{6}\left(1-\frac{1}{4}b\lambda^2\right)\right\}}
\end{eqnarray} 
where we consider only leading order terms in $r_{*}$, i.e. $r_{*}= ln(1-z)^{\frac{1}{3r_+}\left\{1+\frac{\kappa^2\lambda^2}{6}\left(1-\frac{1}{4}b\lambda^2\right)\right\}}$ in obtaining the above expression. We now want to know the function which is independent of $\omega$ and has only $z$ dependence. To do this we first write down eq.(\ref{chs25}) in $z$-coordinate. This reads
\begin{eqnarray}
g(z)\frac{d^{2}A(z)}{dz^2} + g^{\prime}(z)\frac{dA(z)}{dz} &+& \left[\frac{\omega^2}{r^{2}_{+}g(z)}- \frac{2\psi^2(z)}{z^2}\right]A(z)  = 0~.
\label{chs35}
\end{eqnarray}
We now write $A(z)$ as a product of the $\omega$-dependent part and a function of $z$ which we need to determine. Hence, the gauge field reads
\begin{eqnarray}
A(z) = (1-z)^{-\frac{i\omega}{3r_+}\left\{1+\frac{\kappa^2\lambda^2}{6}\left(1-\frac{1}{4}b\lambda^2\right)\right\}} G(z)
\label{chs36}
\end{eqnarray}
where $G(z)$ is regular at the horizon of the black hole.\\
\noindent Substituting this in eq.(\ref{chs35}), we obtain
\begin{eqnarray}
g(z)G^{\prime\prime}(z) + \left[\frac{2i\omega}{3r_+}\left\{ 1+\frac{\kappa^2\lambda^2}{6}\left(1-\frac{1}{4}b\lambda^2 \right) \right\}\frac{g(z)}{1-z} + g^{\prime}(z)\right]G^{\prime}(z) \nonumber \\
\left[\frac{i\omega}{3r_+}\left\{ 1+\frac{\kappa^2\lambda^2}{6}\left(1-\frac{1}{4}b\lambda^2 \right) \right\}\left\{\frac{i\omega}{3r_+} \left\{ 1+\frac{\kappa^2\lambda^2}{6}\left(1-\frac{1}{4}b\lambda^2 \right) \right\}+1 \right\} \frac{g(z)}{(1-z)^2} + \right. \nonumber \\
\left. +\frac{i\omega}{3r_+}\left\{ 1+\frac{\kappa^2\lambda^2}{6}\left(1-\frac{1}{4}b\lambda^2 \right) \right\} \frac{ g^{\prime}(z)}{1-z} +\frac{\omega^2}{r^{2}_{+}g(z)} -\frac{2\psi^2(z)}{z^2}\right]G(z) = 0 ~.~
\label{chs37}
\end{eqnarray}
For $\Delta =1$, we know that 
%\begin{eqnarray}
$\psi(z) = \frac{\langle \mathcal{O}_{1}\rangle}{\sqrt{2} r_{+}} F(z) z$
%\end{eqnarray}
where $F(0) = 1$. For simplification, we consider $F(z)$ to be $1$ because we are neglecting order $\mathcal{O}(z^3)$ term. Substituting this in eq.(\ref{chs37}), we get 
\begin{eqnarray}
3g_{0}(z)G^{\prime\prime}(z) + \left[\frac{2i\omega C_1}{r_+} (1+z+z^2)- 9C_2(z) z^2\right]G^{\prime}(z) +\left[\frac{i\omega C_1}{r_+}\left( 1+z+z^2 -3C_2(z) z^2\right) \right. \nonumber \\
\left. \times \frac{1}{1-z} +\frac{\omega^2}{3r^2_+} \left\{9C_3(z) -(1+z+z^2)^2 C^2_1 \right\}\frac{1}{1-z^3} - \frac{3\langle \mathcal{O}_{1}\rangle^2 }{r^2_{+}}C_4(z) \right]G(z) = 0
\label{chs38}
\end{eqnarray}
where 
\begin{eqnarray}
C_1 = 1 + \frac{\kappa^2\lambda^2}{6}\left(1-\frac{1}{4}b\lambda^2 \right) ~~~&;&~~~ C_2 (z) = \frac{1 +\frac{g^{\prime}_{1}(z)}{g^{\prime}_{0}(z)}}{1+ \frac{g_{1}(z)}{g_{0}(z)}}  \nonumber \\
C_3 (z)= \frac{1}{\left(1+\frac{g_{1}(z)}{g_{0}(z)}\right)^2} ~~~&;&~~~ C_4(z) = \frac{1}{1+\frac{g_{1}(z)}{g_{0}(z)}}~.
\label{chs39}
\end{eqnarray}
Keeping terms upto order $z^3$ in the above equation yields
\begin{eqnarray}
3g_{0}(z)G^{\prime\prime}(z) + \left[\frac{2i\omega C_1}{r_+} (1+z+z^2)- 9C_2(z) z^2\right]G^{\prime}(z) +\left[\frac{i\omega C_1}{r_+}\left\{ 1+2z+3z^2 (1-C_2)  \right. \right. \nonumber \\
\left. \left. + 3z^3 (1-C_2)\right\} +\frac{\omega^2}{3r^2_+} \left\{\frac{C_5 + (C_5-2C^2_1)z +(C_5-5C^2_1)z^2 +(C_5 -7C^2_1)z^3}{1+z+z^2} \right\} \right. \nonumber \\
\left. - \frac{3\langle \mathcal{O}_{1}\rangle^2 }{r^2_{+}} C_4(z) \right]G(z) = 0 ~~~
\label{chs40}
\end{eqnarray}
where $C_5(z) = 9 C_3 (z)- C^2_1$. To solve this equation, we rescale it by letting $z = \frac{z^{\prime}}{a}$, where $a=\frac{<\mathcal{O}_{1}>}{r_+}$ and then take the $a\rightarrow \infty$ limit which corresponds to the low temperature regime \cite{siop}. This leads to 
\begin{eqnarray}
G^{\prime\prime}(z^{\prime}) + G(z^{\prime})= 0~.
\label{chs41}
\end{eqnarray}
The solution of this equation reads
\begin{eqnarray}
G(z^{\prime}) &=& C_+ e^{z^{\prime}} + C_{-}e^{-z^{\prime}} \nonumber \\
\Rightarrow G(z) &=& C_+ e^{a z} + C_{-} e^{-a z} \nonumber \\
&=& C_+ e^{\frac{<\mathcal{O}_{1}>}{r_+}z} + C_{-} e^{-\frac{<\mathcal{O}_{1}>}{r_+}z}~.
\label{chs42}
\end{eqnarray}
The information about the integration constants $C_+$ and $C_{-}$ can be obtained from the appropriate boundary condition. For $\Delta=1$, the boundary condition can be obtained from eq.(\ref{chs40}) by setting $z=1$ in the equation. This gives
\begin{eqnarray}
G^{\prime}(1) + \left[\frac{2\langle \mathcal{O}_{1}\rangle^2 C_4(1)}{r_{+}(3-\frac{2i\omega}{r_+}C_1(1))} -\frac{\frac{i\omega}{3r_+}\left\{3C_1(1)- \frac{i\omega}{9r_+}(4C_5(1)-14C^2_1(1))\right\}}{3-\frac{2i\omega}{r_+}C_1(1)}\right]G(1) =0
\label{chs43}
\end{eqnarray}
where 
\begin{eqnarray}
C_2 (1) =1  ~~~&;& C_3 (1)\approx 1 +\frac{\kappa^2\lambda^2}{3}\left(1-\frac{1}{4}b\lambda^2\right)   \nonumber \\
 C_4 (1) \approx 1 +\frac{\kappa^2\lambda^2}{6}\left(1-\frac{1}{4} b\lambda^2\right)  ~~~&;&~~~ C_5(1) = 9 C_3(1) - C^2_1 ~.
\label{chs44}
\end{eqnarray}
Substituting $G(1)$ and $G^{\prime}(1)$ from eq.(\ref{chs42}) in the boundary condition (\ref{chs43}) yields upto first order in $\omega$
\begin{eqnarray}
\frac{C_+}{C_{-}} = - e^{-2a} \left[\frac{a C_4(1) -3}{a C_4(1) +3} + \frac{2i C_1 \omega}{a r_+}\frac{(2C_4(1) a^2- 3)}{(C_4(1) a + 3)^2} +\mathcal{O}(\omega^2) \right]~.
\label{chs45}
\end{eqnarray}
We finally obtain the solution for $A(z)$ from eqs.(\ref{chs36}), (\ref{chs42}). This reads 
\begin{eqnarray}
A(z) = (1-z)^{-\frac{i\omega}{3r_+}\left\{1+\frac{\kappa^2\lambda^2}{6}\left(1-\frac{1}{4}b\lambda^2\right)\right\}} \left[ C_+ e^{\frac{<\mathcal{O}_{1}>}{r_+}z} + C_{-} e^{-\frac{<\mathcal{O}_{1}>}{r_+}z}\right]~.
\label{chs46} 
\end{eqnarray}
To obtain the conductivity, we now expand the gauge field about $z=0$ :
\begin{eqnarray}
A (z)= A(0) + z A^{\prime}(0) + \mathcal{O}(z^2) ~.
\label{nw6}
\end{eqnarray} 
Now in general $A_{x}$ can be written as
\begin{eqnarray}
A_{x}= A^{(0)}_{x} + \frac{A^{(1)}_{x}}{r_+}z + \mathcal{O}(z^2)~.
\label{nw2}
\end{eqnarray}
Comparing eq.(\ref{nw6}) and eq.(\ref{nw2}), we have  
\begin{eqnarray}
A^{(0)}_{x} = A(0) ~~~;~~~   A^{(1)}_{x} = r_{+} A^{\prime}(0)~.
\end{eqnarray}
Now from the definition of conductivity and gauge/gravity correspondence, we have
\begin{eqnarray}
\label{cmf1}
\sigma (\omega) &=& \frac{\langle J_{x}\rangle}{E_{x}} = \frac{iA^{(1)}_{x}}{\omega A^{(0)}_{x}} =  -\frac{i}{\omega}\frac{r_+ A^{\prime}(z=0)}{A(z=0)}  \nonumber\\ 
&=& \frac{i}{\omega}\frac{1-\frac{C_+}{C_-}}{1+\frac{C_+}{C_-}} \langle\mathcal{O}_1\rangle + \frac{1}{3}\left\{1+\frac{\kappa^2\lambda^2}{6}\left(1-\frac{1}{4}b\lambda^2\right) \right\}~.
\label{chs47}
\end{eqnarray}
Substituting the value of $\frac{C_+}{C_-}$, we obtain the low frequency expression for the conductivity to be
\begin{eqnarray}
\sigma (\omega) = \frac{i\langle\mathcal{O}_1\rangle}{\omega} \left[1+ 2 e^{-2a} \frac{a C_4(1) -3}{a C_4(1) +3} + 4e^{-2a} \frac{i C_1 \omega}{a r_+}\frac{(2C_4(1) a^2- 3)}{(C_4(1) a + 3)^2}  \right] \nonumber \\
+ \frac{1}{3}\left\{1+\frac{\kappa^2\lambda^2}{6}\left(1-\frac{1}{4}b\lambda^2\right) \right\} ~.
\label{chs48}
\end{eqnarray}
A few observations are in order now. This expression for the conductivity is valid in the low temperature limit. The above result is useful in estimating the band gap energy of the holographic superconductors \cite{siop}. This can be estimated as follows. The DC conductivity is defined as the real part of $\sigma$ at $\omega=0$. This reads
%The DC conductivity is now given \textbf{in leading order} by
\begin{eqnarray}
Re~\sigma (\omega= 0) \sim e^{-2a}\left[1+\mathcal{O}(1/a)\right] \approx e^{-2\frac{\langle\mathcal{O}_1\rangle}{r_+}}~. 
\label{chs49}
\end{eqnarray}
Substituting the value of $r_{+}$ in terms of the Hawking temperature $T$ from eq.(\ref{tcm}) in the above equation, we finally obtain
\begin{eqnarray}
Re~\sigma (\omega= 0) \sim e^{-\frac{E_g}{T} }
\end{eqnarray}
where 
\begin{eqnarray}
\label{eg1}
E_g = \frac{3}{2\pi} \left\{1 -\frac{\kappa^2\lambda^2}{6}\left(1-\frac{1}{4}b \lambda^2 \right) \right\} \langle\mathcal{O}_1\rangle.
\end{eqnarray}
$E_g$ is identified to be the band gap energy. Note that the band gap energy gets corrected due to backreaction and the BI parameter. We observe that the effect of the BI parameter vanishes when $\kappa = 0$. We also recover the band gap energy $E_g = \frac{3\langle \mathcal{O}_{1}\rangle}{2\pi} \approx 0.48 \langle \mathcal{O}_{1}\rangle $ \cite{siop} for $\kappa=0$. 
Using the results in Table \ref{t1}, we have calculated the band gap energy for different values of $\kappa$ and $b$. These results are displayed in Table \ref{t2}. We recall that the values of $\lambda^2$ (appearing in eq.(\ref{eg1})) have been estimated using the Sturm-Liouville eigenvalue approach for different values of $\kappa$ \cite{dg1}. The results indicate that for a particular value of the BI parameter $b$, the band gap energy decreases with increasing values of backreaction parameter $\kappa$. Further, for a particular value of $\kappa$, the band gap energy increases with increasing values of the BI parameter $b$. It would be nice to compare our analytical results with numerical studies which are presently missing in the literature to the best of our knowledge.
\begin{table}[h!]
\caption{The values of $\frac{E_g}{\langle\mathcal{O}_{1}\rangle} $ for different values of $\kappa$ and $b$.}
\centering
\begin{tabular}{|c| c| c| c| }
\hline
$\frac{E_{g}}{\langle \mathcal{O}_{1}\rangle} $ & $\kappa =0.1$ & $\kappa =0.2$ & $\kappa = 0.3$  \\
\hline
$b=0.0$ ~& ~0.47646~ & ~0.47344~ & ~0.46845~   \\
\hline
$b=0.1$ & 0.47649 & 0.47356 & 0.46873  \\
\hline
$b=0.2$ & 0.47652 & 0.47369 & 0.46901  \\
\hline
$b=0.3$ & 0.47655 & 0.47382 & 0.46929  \\
\hline
\end{tabular}
\label{t2}
\end{table}

Now we present the self-consistent approach to obtain the conductivity expression. Here we essentially follow the approach in \cite{siop}. We first replace the potential with its average $\langle V \rangle$ in a self-consistent way. With this approximation, the solution of (\ref{chs25}) reads
\begin{eqnarray}
A \sim e^{-i\sqrt{\omega^2 -\langle V\rangle}r_{*}} \sim (1-z)^{-i\sqrt{\omega^2 -\langle V\rangle}\frac{1}{3r_{+}} \left\{1+\frac{\kappa^2\lambda^2}{6}\left(1-\frac{1}{4}b\lambda^2\right)\right\}}~.
\label{chs51}
\end{eqnarray}
%We now expand this solution about $z=0$ to get
%\begin{eqnarray}
%A = 1 + \frac{i}{3} \left\{1+\frac{\kappa^2\lambda^2}{6}\left(1-\frac{1}{4}b\lambda^2\right)\right\}\sqrt{\omega^2-\langle V\rangle} \frac{z}{r_+} + \mathcal{O}(z^2)
%\label{nw11}
%\end{eqnarray}
%Comparing eq.(\ref{nw11}) with eq.(\ref{nw2}), we obtain
%\begin{eqnarray}
%A^{(0)}_{x} = 1, ~~~;~~~ A^{(1)}_{x} = \frac{i}{3} \left\{1+\frac{\kappa^2\lambda^2}{6}\left(1-\frac{1}{8}b\lambda^2\right)\right\}\sqrt{\omega^2 -\langle V\rangle} ~.
%\end{eqnarray}
Once again from the definition of conductivity and gauge/gravity dictionary, we obtain from eq.(\ref{cmf1})
\begin{eqnarray}
\sigma(\omega) &=& -\frac{i}{\omega} \frac{r_{+}A^{\prime}(z=0)}{A(z=0)} \nonumber \\
&=& \frac{1}{3} \left\{1+\frac{\kappa^2\lambda^2}{6}\left(1-\frac{1}{4}b\lambda^2\right)\right\}\sqrt{1-\frac{\langle V\rangle}{\omega^2}} ~.
\label{chs52}
\end{eqnarray}
We now need to estimate the average value of the potential. This reads
\begin{eqnarray}
\langle V\rangle = \frac{\int^{0}_{-\infty} dr_{*} V A^2(r_*)}{\int^{0}_{-\infty} dr_{*} A^2(r_*)}~.
\label{chs53}
\end{eqnarray}
From eq.(\ref{chs26}) and $\psi(z) = \frac{\langle\mathcal{O}_{\Delta}\rangle}{\sqrt{2} r^{\Delta}_{+}} z^{\Delta}F(z)$ the potential reads
\begin{eqnarray}
V \approx \frac{\langle\mathcal{O}_{\Delta}\rangle^{2}}{r^{2\Delta - 2}_{+}} z^{2\Delta -2}\left[g_{0}(z) + g_{1}(z) \right]
\label{chs54}
\end{eqnarray}
where we consider $F(z)\approx 1$. The main contribution to the average value $\langle V \rangle$ in eq.(\ref{chs53}) is from the vicinity of the boundary where $r_{*}\approx -\frac{z}{r_+}$. This is the interesting fact that for the leading order contribution in the nature of $r_{*}$ is independent of $\kappa$ and $b$ parameter. Substituting eq.(\ref{chs54}) in eq.(\ref{chs53}), we obtain the expression for $\langle V \rangle$ to be 
\begin{eqnarray}
\langle V\rangle &\simeq& \frac{\langle\mathcal{O}_{\Delta}\rangle^{2}}{r^{2\Delta - 2}_{+}}\left\{ \frac{\int^{\infty}_{0} dz e^{2i\sqrt{\omega^2 -\langle V\rangle}\frac{z}{r_+}} z^{2\Delta -2} g_{0}(z)}{\int^{\infty}_{0} dz e^{2i\sqrt{\omega^2 -\langle V\rangle}\frac{z}{r_+}}} \right\} \nonumber \\
%- \frac{\int^{\infty}_{0} dz e^{2i\sqrt{\omega^2 -<V>}\frac{z}{r_+}} z^{2\Delta -2} g_{1}(z)}{\int^{\infty}_{0} dz e^{2i\sqrt{\omega^2 -<V>}\frac{z}{r_+}}}\right\} 
&\simeq & \langle\mathcal{O}_{\Delta}\rangle^{2} \left[ \frac{\Gamma(2\Delta -1)}{(-2i\sqrt{\omega^2 - \langle V\rangle})^{2\Delta -2}} \right] ~.
%-\frac{r^3_+ \Gamma(2\Delta +2)}{(-2i\sqrt{\omega^2 - <V>})^{2\Delta +1}} + \frac{\kappa^2 \lambda^2}{2}\left\{ \frac{r^4_+ \Gamma(2\Delta +3)}{(-2i\sqrt{\omega^2 - <V>})^{2\Delta +2}} \right.\right. \nonumber \\
%\left. \left. - \frac{r^3_+ \Gamma(2\Delta +2)}{(-2i\sqrt{\omega^2 - <V>})^{2\Delta +1}}  -\frac{b\lambda^2}{40} \left(\frac{r^8_+\Gamma(2\Delta +7)}{(-2i\sqrt{\omega^2 - <V>})^{2\Delta +6}} -\frac{r^3_+ \Gamma(2\Delta +2)}{(-2i\sqrt{\omega^2 - <V>})^{2\Delta +1}}  \right)\right\} \right]
\label{chs55}
\end{eqnarray}
This is the self consistent equation for the average value of the potential $\langle V\rangle$, which depends on the frequency $\omega$.
At the low frequency limit, we set $\omega =0$ in eq.(\ref{chs55}) which leads to
\begin{eqnarray}
\langle V\rangle^{\Delta} = \frac{\langle\mathcal{O}_{\Delta}\rangle^{2}}{2^{2\Delta-2}} \Gamma(2\Delta-1) ~.
%-\frac{r^3_+ \Gamma(2\Delta+2)}{2^3 <V>^3} +\frac{\kappa^2\lambda^2}{2}\left\{ \frac{r^4_+ \Gamma(2\Delta+3)}{2^4 <V>^4} \right. \right. \nonumber \\
%\left. \left. - \frac{r^3_+ \Gamma(2\Delta+2)}{2^3 <V>^3} -\frac{b\lambda^2}{40} \left( \frac{r^8_+ \Gamma(2\Delta+7)}{2^8 <V>^8}- \frac{r^3_+ \Gamma(2\Delta+2)}{2^3 <V>^3}\right) \right\} \right]
\label{chs56}
\end{eqnarray}
For $\Delta=1$, this gives
\begin{eqnarray}
\langle V\rangle = \langle\mathcal{O}_{1}\rangle^{2} ~.
%\left[\Gamma(1) -\frac{r^3_+ \Gamma(4)}{2^3 \langle V\rangle^3} +\frac{\kappa^2\lambda^2}{2}\left\{ \frac{r^4_+ \Gamma(5)}{2^4 <V>^4} \right. \right. \nonumber \\
%\left. \left. - \frac{r^3_+ \Gamma(4)}{2^3 <V>^3} -\frac{b\lambda^2}{40} \left( \frac{r^8_+ \Gamma(9)}{2^8 <V>^8}- \frac{r^3_+ \Gamma(4)}{2^3 <V>^3}\right) \right\} \right]
\label{chs57}
\end{eqnarray}
Using this in eq.(\ref{chs52}), the conductivity is given by
\begin{eqnarray}
\sigma (\omega) &=& \frac{1}{3} \left\{1+\frac{\kappa^2\lambda^2}{6}\left(1-\frac{1}{4}b\lambda^2\right)\right\}\sqrt{1-\frac{\langle\mathcal{O}_{1}\rangle^2}{\omega^2}} \nonumber \\
&=& \frac{i\langle\mathcal{O}_{1}\rangle }{3\omega}\left\{1+\frac{\kappa^2\lambda^2}{6}\left(1-\frac{1}{4}b\lambda^2\right)\right\} \sqrt{1-\frac{\omega^2}{\langle\mathcal{O}_{1}\rangle^2}}~.
\label{chs60}
\end{eqnarray}
It can be observed that the above result agrees in form at the leading order with the result obtained in eq.(\ref{chs48}). This feature was also observed in \cite{siop}. However, this method does not capture the expression for the band gap energy.

\section{Conclusions}
We have analytically computed the conductivity of holographic superconductors in the framework of Born-Infeld electrodynamics away from the probe limit. By employing a perturbative approach, we have computed the backreacted bulk spacetime metric taking into account the effect of the Born-Infeld electrodynamics. We then moved onto compute the conductivity which is found to contain the effects of the backreaction parameter $\kappa$ and the BI parameter $b$. From the real part of the conductivity (computed at $\omega=0$), the band gap energy is obtained. It is observed that the energy gap gets corrected from the standard value due to the parameters $\kappa$ and $b$. The dependence of the band gap energy on the non-linear effects coming from BI electrodynamics is manifest. The results show that the band gap energy decreases with increase in the values of the backreaction parameter $\kappa$ for a fixed value of the BI parameter $b$. Moreover, it increases with increase in $b$ for a fixed value of $\kappa$. We then perform the computation of conductivity by following a self consistent approach and finally compare the results obtained from the two approaches. As a future work, we can extend our analysis for Gauss-Bonnet black holes in $5$-dimensions. Work in this direction is in progress.

\section*{Acknowledgments} DG would like to thank DST-INSPIRE, Govt. of India for financial support and to thank IUCAA, Pune for providing wonderful hospitality during IUCAA visit. SG acknowledges the support by DST SERB under Start Up Research Grant (Young Scientist), File No.YSS/2014/000180. SG also acknowledge the Visiting Associateship at IUCAA during which a substantial part of this work was completed. The authors thank the referee for useful comments.

%%%%%%%%%%%%%%%%%%%%%%%%%%%%%%%%%%%%%%%%%%%%%%%%%%

\end{document}